\documentclass[aps,prl,superscriptaddress,twocolumn]{revtex4-1}

\usepackage{graphics}
\usepackage{graphicx}
\usepackage{dcolumn}
\usepackage{bm}

\begin{document}


\title{Magnetically driven metal-insulator transition in NaOsO$_3$}



\author{S.~Calder}
\email{caldersa@ornl.gov}
\affiliation{Oak Ridge National Laboratory, Oak Ridge, Tennessee 37831, USA.}

\author{V.~O.~Garlea}
\affiliation{Oak Ridge National Laboratory, Oak Ridge, Tennessee 37831, USA.}

\author{D.~F.~McMorrow}
\affiliation{London Centre for Nanotechnology, University College London, London, WC1H 0AH, UK.}

\author{M.~D.~Lumsden}
\affiliation{Oak Ridge National Laboratory, Oak Ridge, Tennessee 37831, USA.}

\author{M.~B.~Stone}
\affiliation{Oak Ridge National Laboratory, Oak Ridge, Tennessee 37831, USA.}

\author{J.~C.~Lang}
\affiliation{Advanced Photon Source, Argonne National Laboratory, Argonne, Illinois 60439, USA.}

\author{J.-W.~Kim}
\affiliation{Advanced Photon Source, Argonne National Laboratory, Argonne, Illinois 60439, USA.}

\author{J.~A.~Schlueter}
\affiliation{Materials Science Division, Argonne National Laboratory, Argonne, Illinois 60439, USA.}

\author{Y.~G.~Shi}
\affiliation{Institute of Physics, Chinese Academy of Sciences, 100190 Beijing, China.}
\affiliation{Superconducting Properties Unit, National Institute for Materials Science,1-1 Namiki, Tsukuba, 305-0044 Ibaraki, Japan.}

\author{K.~Yamaura}
\affiliation{Superconducting Properties Unit, National Institute for Materials Science,1-1 Namiki, Tsukuba, 305-0044 Ibaraki, Japan.}
\affiliation{JST, Transformative Research-Project on Iron Pnictides (TRIP), 1-1 Namiki, Tsukuba, 305-0044 Ibaraki, Japan.}

\author{Y.~S.~Sun}
\affiliation{International Center for Materials Nanoarchitectonics (MANA), National Institute for Materials Science, Tsukuba, Ibaraki 305-0044, Japan.}

\author{Y. Tsujimoto}
\affiliation{International Center for Materials Nanoarchitectonics (MANA), National Institute for Materials Science, Tsukuba, Ibaraki 305-0044, Japan.}

\author{A.~D.~Christianson}
\affiliation{Oak Ridge National Laboratory, Oak Ridge, Tennessee 37831, USA.}


\date{\today}

\begin{abstract}
The metal-insulator transition (MIT) is one of the most dramatic manifestations of electron correlations in materials. Various mechanisms producing MITs have been extensively considered, including the Mott (electron localization via Coulomb repulsion), Anderson (localization via disorder) and Peierls (localization via distortion of a periodic 1D lattice). One additional route to a MIT proposed by Slater, in which long-range magnetic order in a three dimensional system drives the MIT, has received relatively little attention. Using neutron and X-ray scattering we show that the MIT in NaOsO$_3$ is coincident with the onset of long-range commensurate three dimensional magnetic order. Whilst candidate materials have been suggested, our experimental methodology allows the first definitive demonstration of the long predicted Slater MIT. We discuss our results in the light of recent reports of a Mott spin-orbit insulating state in other $5d$ oxides.
\end{abstract}

\maketitle

The precise microscopic origin behind metal-insulator transitions (MIT) has been one of the enduring problems within condensed matter physics  \cite{ImadaMIT}. Many materials exhibit a Mott MIT that is explained by invoking strong Coulomb interactions ($U$) which open a gap at the Fermi energy forcing the conductor into an insulating phase \cite{NFMott,JHubbard}. A Mott MIT is independent of magnetic correlations. One alternative route to a MIT, known as the Slater transition, is driven by antiferromagnetic (AFM) order alone opening a band gap independent of strong $U$ \cite{Slater}. Although magnetism is often present in the phase diagram of a MIT, there have been no definitive experimental examples of a three dimensional (3D) magnetically driven Slater MIT.  

The $5d$ transition metal oxide NaOsO$_3$, along with Cd$_2$Os$_2$O$_7$ and $Ln_2$Ir$_2$O$_7$ ($Ln$ = Lanthanide),  have been suggested on the basis of bulk measurements to host a Slater MIT \cite{ShiPRB, Sleight1974357,MandrusCd2Os2O7, PhysRevB.66.035120, JPSJ.80.094701,PhysRevB.83.180402}. However, to define a material as undergoing a Slater MIT it is obligatory to show microscopic 3D commensurate long-range magnetic ordering concurrent with the MIT. There exists no such experimental evidence in any candidate material. For Cd$_2$Os$_2$O$_7$ and $Ln_2$Ir$_2$O$_7$ the elusiveness of experimental verification can be attributed to the inherent magnetic frustration in the pyrochlore lattice and prohibitively high neutron absorption values for Cd and Ir. We have investigated the continuous MIT in NaOsO$_3$ with neutron and X-ray scattering, observed long-range magnetic order at the MIT, and determined the nature of that order on a microscopic level. Thus presenting the first definitive experimental example of a Slater MIT. 

\begin{figure}[b]
     \centering
                  \includegraphics[width=0.65\columnwidth]{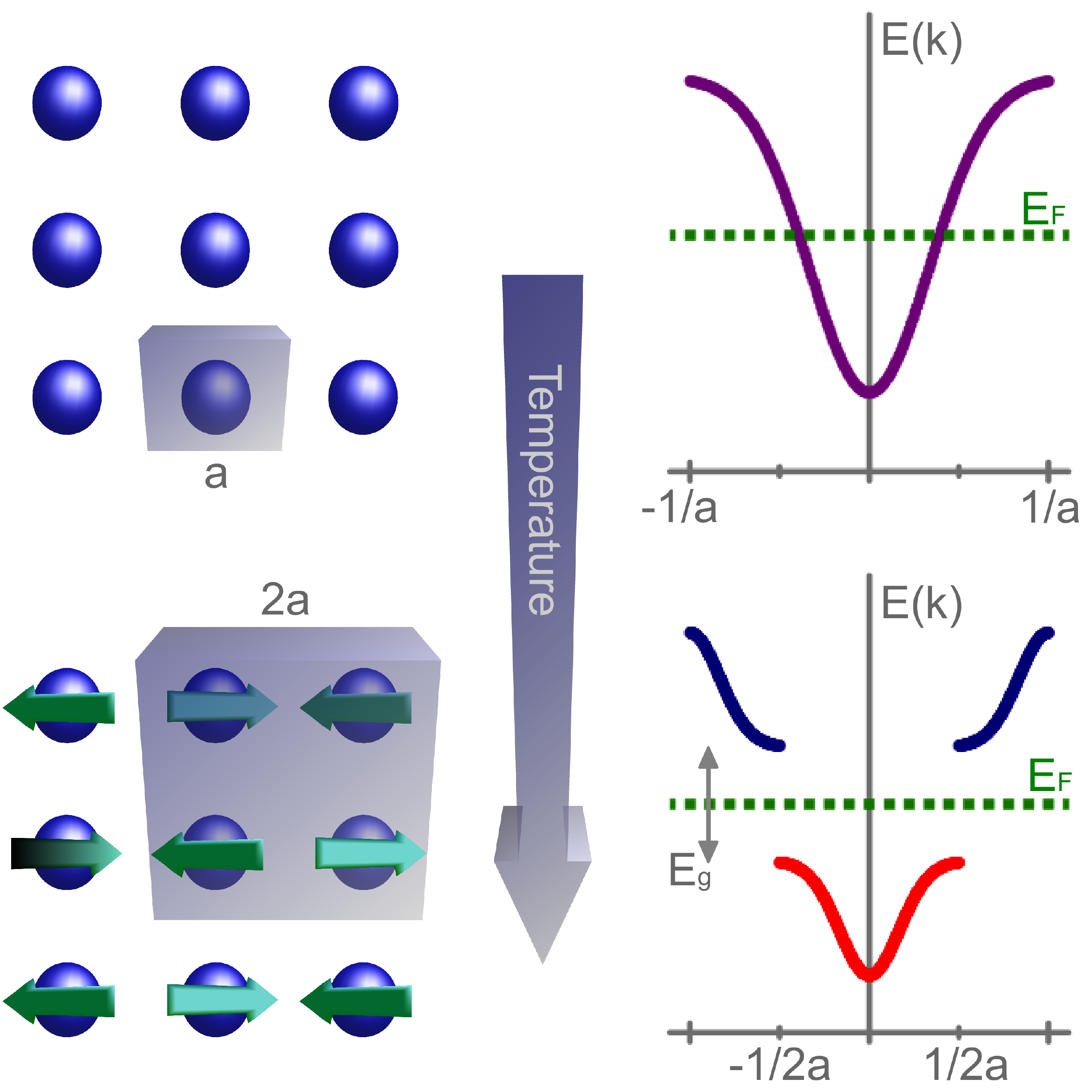}
 \caption{\label{Slater_schematic}Slater MIT: (Left) commensurate AFM order doubles the unit cell. (Right) The opposite potential on each neighboring ion in the magnetic regime splits the electronic band, creating an insulating band gap, E$_g$.}
\end{figure}

Slater's description of a magnetically driven MIT is shown schematically in Fig.~\ref{Slater_schematic}. Commensurate AFM order, with every neighboring spin oppositely aligned, occurs at the MIT temperature (T$_{\rm MIT}$) creating an opposite periodic potential on each nearest neighbor. This results in an energy gap that splits the electronic band at the newly created magnetic Brillouin zone boundary. In a half-filled electronic outer-level the lower band will be preferentially fully occupied and the upper band empty. The emergence of this band gap causes the system to undergo a Slater MIT, and provides a canonical example of an itinerant system that can be described by the self-consistent single electron Hartree-Fock theory \cite{FGebhard}.

Single crystal (0.2$\times$0.2$\times$0.05mm) and polycrystalline (4.6g) samples of NaOsO$_3$ were prepared in pressures up to  6 GPa as described in Ref.~\cite{ShiPRB}. Neutron powder diffraction (NPD) was performed at the High Flux Isotope Reactor (HFIR) using beamline HB-2A with a  wavelength of $\lambda$=$1.54$ $\rm \AA$. The results were Rietveld refined with Fullprof and representational analysis performed using SARA$h$ \cite{sarahwills}. Polarized and unpolarized elastic measurements were performed on the spectrometer HB-1 at HFIR with $\lambda$$=$$2.46$ $\rm \AA$. Polarization was performed using a vertical focusing Heussler monochromator with a vertical guide field with flipping ratio 14. Magnetic resonant X-ray scattering (MRXS) was carried out at beamline 6-ID-B at the Advanced Photon Source (APS) in reflection mode. Graphite was used as the polarization analyzer crystal at the (0,0,10) and (0,0,8) reflections on the L2 and L3 edges, respectively, to achieve the analyzer scattering angle close to 90$^\circ$. To account for absorption, energy scans were performed without the analyzer and with the detector away from any Bragg peaks  to observe the fluorescence signal at both edges.

A first-order structural transition is often associated with a Mott MIT, however this should not occur in the continuous Slater MIT. To test this NPD was performed through T$_{\rm  MIT} \approx$  410 K of NaOsO$_3$ from 200 K to 500 K. The results were refined to a structural model of a distorted perovskite structure with space group $Pnma$, in agreement with results in Ref.~\cite{ShiPRB}. We note that $\chi^2$ remained consistent for all temperatures with a range of $2.29 \leq \chi^2 \leq 3.29$ (see supplemental material for a complete set of refinement parameters). Taken along with the smooth thermal shifting of Bragg peaks observed strongly indicates that the $Pnma$ structural model is equally applicable, and correct, above and below T$_{\rm MIT}$. 

\begin{figure}[tb]
     \centering
                  \includegraphics[width=1.0\columnwidth]{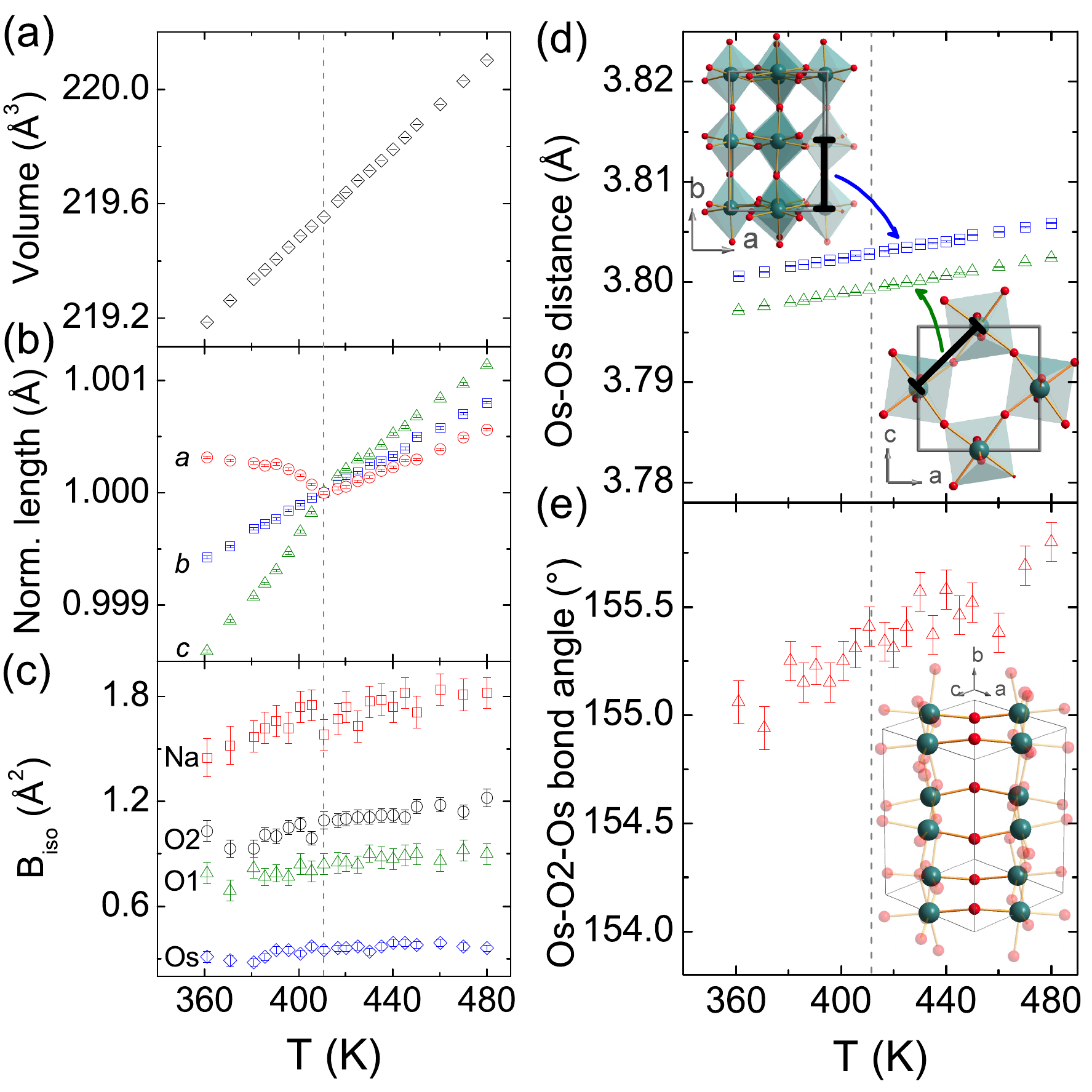}
 \caption{\label{Figure2Calder}Refined structural parameters from NPD. The vertical dashed line indicates T$_{\rm MIT}$. The insets of (d) and (e) is a representation of the crystallographic bond lengths and angles considered, with the larger spheres indicating the Os ions and the small spheres indicating the oxygen ions. The crystal structure shows an anomaly in $a$ and $c$ lattice constants at T$_{\rm MIT}$. However, there is no indication of a structural symmetry change.}
\end{figure}

The temperature variation of the unit cell parameters from NPD for NaOsO$_3$ are shown in Fig.~\ref{Figure2Calder}. The unit cell volume shows no deviation through T$_{\rm MIT}$ with the small expansion of under $0.1 \%$ indicative of a stable crystal structure, Fig.~\ref{Figure2Calder}(a). The anomaly observed in the $a$ and $c$ axis at T$_{\rm MIT}$ (Fig.~\ref{Figure2Calder}(b)) is not an indication of a structural symmetry change or unstable structure, as evidenced by the stable thermal parameters  through T$_{\rm MIT}$ in Fig.~\ref{Figure2Calder}(c). Instead the $a$-$c$ structural behavior can be understood by considering the ions that control the lattice constants, namely the Os-Os bond distance (Fig.~\ref{Figure2Calder}(d)) and Os-O2-Os bond angle (Fig.~\ref{Figure2Calder}(e)). To relieve tension and reduce energy the distorted octahedra prefer to align with the unit cell axis as the temperature is lowered, resulting in the Os-O2-Os bond angle decreasing. This bond is overwhelmingly along the $a$ and $c$ directions and as such a change in Os-O2-Os bond angle results in a change in $a$ and $c$ unit cell lengths, as observed. Consequently the Os ions along the $a$-$c$ axis are pulled closer together. When sufficiently close, the Os ions interact and form the magnetic long range ordered phase observed. This suggests a magnetostriction control parameter in which it would be possible to influence T$_{\rm MIT}$ by controlling the $a$-$c$ unit cell lengths with the application of pressure.

\begin{figure}[tb]
     \centering
                  \includegraphics[width=1.0\columnwidth]{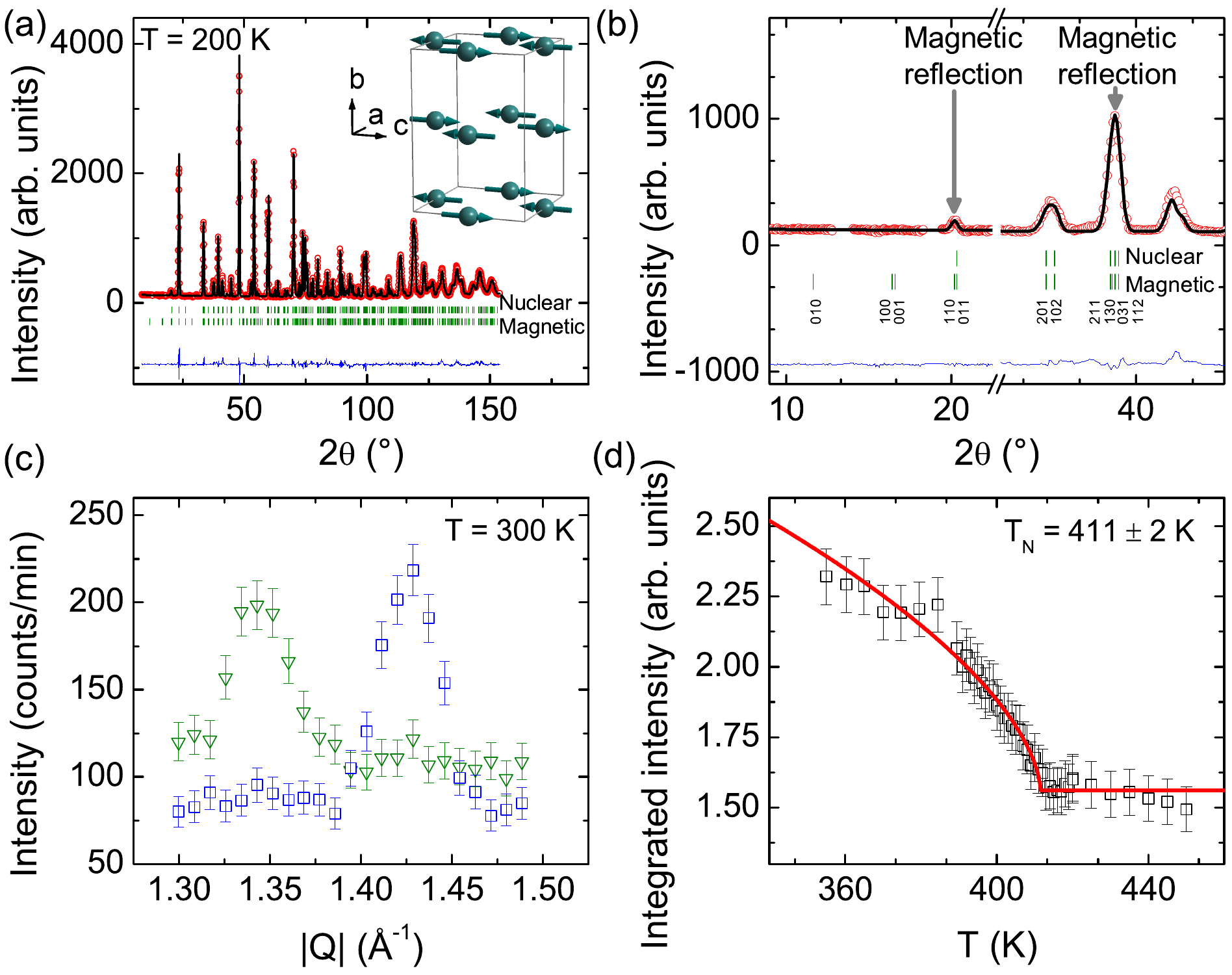}
 \caption{\label{Figure3Calder}(a)-(b) NPD modeled with crystallographic space group $Pnma$ and G-type AFM ordering. Magnetic order is shown schematically inset, with only the Os ions represented for clarity.~(c) Polarized NPD with polarization parallel to the scattering vector with flipper-on (triangles) and flipper-off (squares) unambiguously assigns the scattering centered around $|Q|$=1.43 $\rm \AA^{-1}$ as purely magnetic.  The lower $|Q|$$\approx$1.35 $\rm \AA^{-1}$ structural peak corresponds to an unknown non-magnetic impurity phase.~(d) Integrated intensity of the $|Q|$=1.43 $\rm \AA^{-1}$ peak as a function of temperature.  Solid line is a power law fit as described in the text.}
\end{figure}

Additional Bragg peaks develop below T$_{\rm MIT}$ in the NPD pattern in Fig.~\ref{Figure3Calder}(a)-(b), indicative of magnetic order. Measuring the commensurate peak at $|Q|$$\approx$$1.43$ $\rm \AA^{-1}$, consistent with the (110) and (011) structural reflections, with polarized neutrons unambiguously assigns the additional scattering below T$_{\rm MIT}$ as arising from magnetic order, see Fig.~\ref{Figure3Calder}(c). To determine the nature of the magnetic order we implemented representational analysis \cite{sarahwills}. For a second order transition, Landau theory states that the symmetry properties of the magnetic structure are described by only one irreducible representation. For the $Pnma$ crystal structure with the magnetic moment on the Os ion and commensurate propagation vector, ${\bf k} = (000)$, there are four possible irreducible representations. Only one gave a correct description of the magnetic scattering intensities in NaOsO$_3$, labelled $\Gamma(5)$, with only scattering from spins along the c-axis producing intensities at the correct reflections, as shown in Fig.~\ref{Figure3Calder}(a)-(b). 

We have thus  established NaOsO$_3$ to have AFM long range order, with the spins oriented along the $c$-axis in the G-type AFM structure, shown schematically in Fig.~\ref{Figure3Calder}(a). The magnetic moment found from the refinement is 1.0(1)$ \mu_{\rm B}$. A reduced moment is suggestive of the itinerant nature of the $5d$ magnetic ions resulting in a large degree of covalency. A survey of specific Bragg peaks  (polarized and unpolarized) through the magnetic transition revealed no evidence for canting of the spins along either the $a$ or $b$-axis that would be indicated by additional magnetic Bragg peaks compatible with the $\Gamma(5)$ irreducible representation. Figure~\ref{Figure3Calder}(d) establishes the AFM transition temperature from the integrated intensity of unpolarized neutron scattering around $|Q|$$\approx$$1.43$ $\rm \AA^{-1}$ to be T$_{\rm N}$$= $$411.(2)$ K with $\beta$$\approx$$0.3(1)$. The scattering is best described as 3D and does not have any of the features necessary to be 2D. Therefore AFM order occurs at the same temperature as the change in resistivity thus linking the AFM-MIT transitions at T$_{\rm MIT}$.

\begin{figure}[tb]
     \centering
\includegraphics[width=0.9\columnwidth]{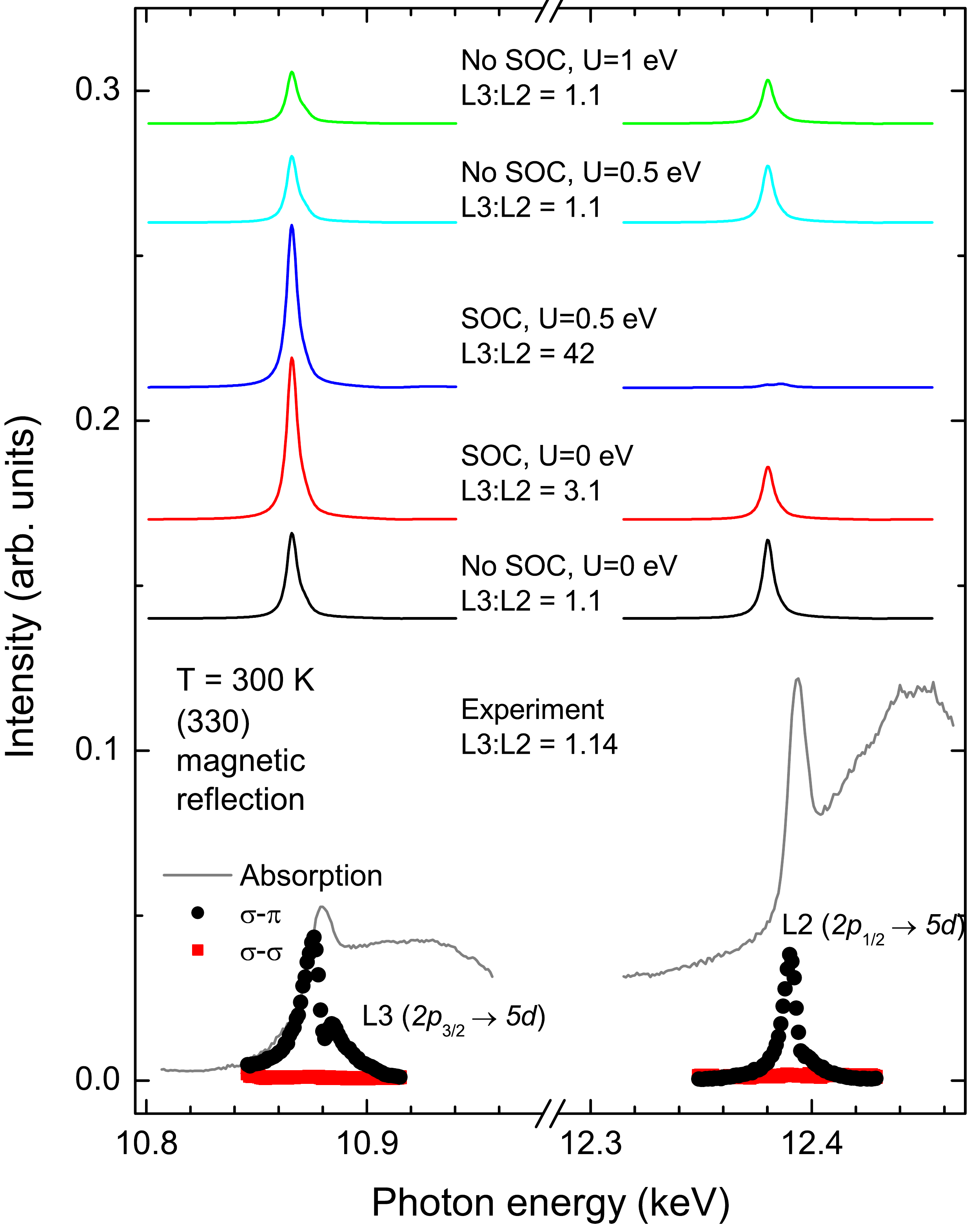}
 \caption{\label{Figure4Calder}Measurement of the resonant enhancement at the L2 and L3  edges. $\sigma$ ($\pi$) corresponds to polarization perpendicular (parallel) to the scattering plane. The $\sigma$-$\pi$ scattering, sensitive to the resonant magnetic scattering, is evident at the expected resonant energies, with a similar enhancement of $\sim$$10^2$ at the L2 and L3 edges. In the $\sigma$-$\sigma$ channel no scattering is observed, showing that the resonant intensity is $\pi$ polarized, indicating the resonant enhancement is purely from magnetic scattering. The 5 pairs of solid lines in the top portion of the figure are results from FDMNES calculations for specified SOC and $U$. The ratio of L3:L2 intensities is noted for each of these calculations.}
\end{figure}

We investigated the role of spin-orbit coupling (SOC) and $U$ by probing the Os $5d^3$ electrons in a single crystal of NaOsO$_3$ using MRXS. MRXS selectively measures the enhanced signal at an X-ray absorption edge, allowing a direct probe of the $t_{2g}$ $5d$ electrons. Measurements at both the L2 edge, corresponding to  a $2p_{\frac{1}{2}} \rightarrow 5d$ transition,  and L3 edge ($2p_{\frac{3}{2}} \rightarrow 5d$) in Os were undertaken at the (330) and (550) reflections, Fig.~\ref{Figure4Calder} (see supplemental material for measurements through T$_{\rm MIT}$ that support the NPD behavior). Significantly the resonant enhancement is approximately the same at both the L2 and L3 edges. A similar ratio of the resonant scattering was observed for K$_2$ReCl$_6$, in which Re$^{4+}$ has the same $5d^3$ electronic configuration as Os$^{5+}$\cite{0953-8984-15-2-108}. Whereas the resonant enhancement for Sr$_2$IrO$_4$ at the L2 edge had only $\sim$$1 \%$ of the intensity observed at the L3 edge, allowing MRXS results to directly show the enhanced role of SOC \cite{KimScience}. Compared to the more extensively studied $3d$ systems, $5d$ transition metal oxides are expected to have less localized electrons with intrinsically weaker Coulomb interactions due to the increase in the radius of the electronic wavefunction. Conversely, the intra-atomic SOC, normally only considered as a perturbation in $3d$ systems, increases. In Sr$_2$IrO$_4$ SOC breaks the degeneracy of the crystal-field separated $t_{2g}$ manifold into a fully occupied $J_{\rm \it eff}=\frac{3}{2}$ quadruplet and a singly occupied $J_{\rm \it eff}=\frac{1}{2}$ doublet with $L_{\rm \it eff} = 1$, creating a Mott spin-orbit state \cite{KimScience,NaturePesin}. SOC similarly dominates in BaIrO$_3$  \cite{PhysRevLett.105.216407}, an isostructural material to NaOsO$_3$. While Ir$^{4+}$ has a $5d^5$ electronic configuration, Os$^{5+}$ in NaOsO$_3$ has $5d^3$. Consequently, even for large SOC, Hund's coupling due to preserving the degeneracy of occupying each $t_{2g}$ orbital ($d_{xy}$, $d_{yz}$ and $d_{zx}$) is predicted to dominate \cite{ChenPRBd2}. We present calculations to reproduce the MRXS scattering using the FDMNES program \cite{FDMNESJoly}. This takes into account SOC and $U$ and has been shown to succesfully reproduce MRXS results for $5d$ systems \cite{FDMNESboseggia}. We ran this {\it ab initio} program using fully relativistic monoelectronic DFT-LSDA calculations on the basis of Green's formalism for a muffin-tin potential for a cluster radius of 3.2 $\rm \AA$, results are shown as pairs of solid lines at the top of Fig. \ref{Figure4Calder}. The experimental resonant scattering ratio of edges L3:L2 is most closely reproduced for the case of no SOC and $U = 0$. The addition of SOC has the effect of significantly increasing the ratio of L3:L2 that is not reflected in the experimental results, indicating SOC does not alter the electronic configuration in NaOsO$_3$. Therefore, contrary to the recent results for Ir$^{4+}$ in which the effect of SOC is comparably large, in NaOsO$_3$ SOC does not play a significant role in the creation of the magnetic insulating phase. Introducing Coulomb interactions through $U$ decreases the agreement with the experimental MRXS L2-edge shape as $U$ increases, this concurs with what would be expected of a Slater MIT in which Coulomb interactions are negligible. 

The neutron and X-ray experimental results presented have all the required elements to describe NaOsO$_3$ as undergoing a Slater MIT. We have shown that long-range commensurate AFM order occurs at the same temperature as the continuous MIT. The AFM order is G-type, thus there is the necessary opposite potential surrounding each Os ion in 3D that creates an electronic band gap at T$_{\rm MIT}$. Due to the magnetic Os$^{5+}$ ion having a half-full $t_{2g}$ outer-level ($5d^3$) the development of this band gap results in a continuous MIT. The $t_{2g}$ degeneracy of single occupied $d_{xy}$, $d_{yz}$ and $d_{zx}$ orbitals precludes a structural change through Jahn-Teller distortions. We observed no distortion of the O-Os bond distances and no other evidence for a symmetry change through T$_{\rm MIT}$. We find no indication that the large SOC plays a significant role in the magnetic insulating state and we conclude that SOC does not alter the electronic configuration and break the degeneracy of the $t_{2g}$ manifold. Instead a reduced Os moment of $\sim$1$\mu_B$ is observed from our NPD results that can be explained in an itinerant model as being a consequence of the increased hybridization between the Os $d$ orbitals and the oxygen $p$ orbitals. As an intriguing comparison a reduced moment and extremely high magnetic ordering temperature of over 1000 K is observed in the $4d$ perovskite SrTcO$_3$ \cite{PhysRevLett.106.067201}, a system that has the same $d^3$ electron configuration with electron hybridization and strikingly similar magnetic properties to NaOsO$_3$.

The Slater MIT is based on single electron itinerant physics. This contrasts to Mott physics in which strong Coulomb interactions play a central role. This is a key distinction between a magnetically driven MIT being described as a Slater transition or magnetism being simply observed to occur at the Coulomb driven MIT, as is the case in a Mott-Heisenberg transition \cite{FGebhard}. Often systems are found to lie in an intermediate regime between Mott (local moment) and Slater (itinerant), and indeed this is an open question in unconventional superconductors. There has been recent debate as to the role of SOC on the electron configurations in $5d$ transition metal oxides. Both the $5d^5$ Ir and $5d^3$ Os have large SOC and strong crystal-fields. However, the SOC alters the $t_{2g}$ degeneracy in Sr$_2$IrO$_4$ and BaIrO$_3$ leading to local moment behavior. Conversely for NaOsO$_3$ the different number of electrons does not favor the breaking of the $t_{2g}$ manifold by the strong SOC, resulting in the degeneracy of the half-filled $t_{2g}$ orbitals remaining to conserve Hund's rules. Therefore the $5d$ osmium ion in NaOsO$_3$ results in the expected itinerant behavior that, coupled with the long-range G-type AFM structure that we have revealed, allows NaOsO$_3$ to host a continuous 3D Slater MIT above room temperature.

\begin{acknowledgments}
Work at ORNL was supported by the scientific User Facilities Division, Office of Basic Energy Sciences, U.~S.~Department of Energy (DOE). Use of the Advanced Photon Source, an Office of Science User Facility operated for the U.S. DOE Office of Science by Argonne National Laboratory, was supported by the U.S. DOE under Contract No. DE-AC02-06CH11357.  This research was supported in part by the Grants-in-Aid for Scientific Research (22246083) from JSPS; and the Funding Program for World-Leading Innovative R$\rm \&$D on Science and Technology (FIRST Program) from JSPS.
\end{acknowledgments}


\begin{thebibliography}{10}

\bibitem{ImadaMIT}
Imada, M., Fujimori, A., and Tokura, Y.
\newblock {\em Rev. Mod. Phys.}{ \bf 70,} 1039--1263 (1998).

\bibitem{NFMott}
Mott, N.~F.
\newblock {\em Proc. Phys. Soc. Lond. A}{ \bf A62,} 416–422 (1949).

\bibitem{JHubbard}
Hubbard, J.
\newblock {\em Proc. R. Soc.}{ \bf A276,} 238 (1963).

\bibitem{Slater}
Slater, J.~C.
\newblock {\em Phys. Rev.}{ \bf 82,} 538--541 (1951).

\bibitem{ShiPRB}
Shi, Y.~G., et al.
\newblock {\em Phys. Rev. B}{ \bf 80,} 161104 (2009).


\bibitem{Sleight1974357}
Sleight, A.~W., Gillson, J.~L., Weiher, J.~F., and Bindloss, W.
\newblock {\em Solid State Comm.}{ \bf 14,} 357--359 (1974).

\bibitem{MandrusCd2Os2O7}
Mandrus, D., et al.
\newblock {\em Phys. Rev. B}{ \bf 63,} 195104 (2001).


\bibitem{PhysRevB.66.035120}
Padilla, W.~J., Mandrus, D., and Basov, D.~N.
\newblock {\em Phys. Rev. B}{ \bf 66,} 035120 (2002).

\bibitem{JPSJ.80.094701}
Matsuhira, K., Wakeshima, M., Hinatsu, Y., and Takagi, S.
\newblock {\em J. Phys. Soc. Jpn.}{ \bf 80,} 094701 (2011).

\bibitem{PhysRevB.83.180402}
Zhao, S., et al.
\newblock {\em Phys. Rev. B}{ \bf 83,} 180402 (2011).

\bibitem{FGebhard}
Gebhard, F.
\newblock {\em {\it The {M}ott {M}etal-{I}nsulator {T}ransition}}.
\newblock Springer, Berlin,  (1997).

\bibitem{sarahwills}
Wills, A.
\newblock {\em Physica B}{ \bf 276,} 680--681 (2000).

\bibitem{0953-8984-15-2-108}
McMorrow, D.~F., Nagler, S.~E., McEwen, K.~A., and Brown, S.~D.
\newblock {\em J. Phys.: Condens. Matter}{ \bf 15,} L59 (2003).

\bibitem{KimScience}
Kim, B.~J., et al.
\newblock {\em Science}{ \bf 323,} 1329--1332 (2009).

\bibitem{NaturePesin}
Pesin, D. and Balents, L.
\newblock {\em Nature Phys.}{ \bf 6,} 376 (2010).

\bibitem{PhysRevLett.105.216407}
Laguna-Marco, et al.
\newblock {\em Phys. Rev. Lett.}{ \bf 105,} 216407 (2010).

\bibitem{ChenPRBd2}
Chen, G.~ and Balents, L.~ et al.
\newblock {\em Phys. Rev. B}{ \bf 84,} 094420 (2011).


\bibitem{FDMNESJoly}
Joly, Y.
\newblock {\em Phys. Rev. B}{ \bf 84,} 094420 (2001).

\bibitem{ FDMNESboseggia}
S.~Boseggia, R.~Springell, H.~C.~Walker, A.~T.~Boothroyd, D.~Prabhakaran, D.~Wermeille, L.~Bouchenoire, S.~P.~Collins, and D.~F.~McMorrow
\newblock {\em arXiv:1201.1452v1} (2012).

\bibitem{PhysRevLett.106.067201}
Rodriguez, Efrain E, et al.
\newblock {\em Phys. Rev. Lett.}{ \bf 106,} 067201 (2011).


\end{thebibliography}
\end{document}